\documentclass[10pt,conference]{IEEEtran}

\usepackage{cite}
\usepackage{graphicx}
\usepackage{psfrag}
\usepackage{subfigure}
\usepackage{flushend}
\usepackage{url}
\usepackage{color}

\usepackage{amsmath}
\usepackage{array}
\usepackage{graphics}
\usepackage{epsfig}
\usepackage{amsbsy}
\usepackage{amssymb}
\usepackage{amsthm}
\usepackage{amsmath}

\usepackage{framed}
\usepackage{algorithm}
\usepackage{algorithmic}
\newtheorem{proposition}{Proposition} 

\IEEEoverridecommandlockouts

\begin{document}

\title{On Information and Energy Cooperation in\\ Energy Harvesting Cognitive Radio}
\author{
\IEEEauthorblockN{Jeya~Pradha~J, Sanket~S.~Kalamkar\IEEEauthorrefmark{1}, and~Adrish~Banerjee}
\IEEEauthorblockA{Department of Electrical Engineering, Indian Institute of Technology Kanpur, 208016, India\\
Email: \{pradha, kalamkar, adrish\}@iitk.ac.in}
\thanks{\IEEEauthorrefmark{1}The author is supported by the Tata Consultancy Services (TCS) research fellowship.}
}

\maketitle

\begin{abstract}
This paper considers the cooperation between primary and secondary users at information and energy levels when both users are energy harvesting nodes. In particular, a secondary transmitter helps relaying the primary message, and in turn, gains the spectrum access as a reward. Also, the primary transmitter supplies energy to the secondary transmitter if the latter is energy-constrained, which facilitates an uninterrupted cooperation. We address this two-level cooperation over a finite horizon with the finite battery constraint at the secondary transmitter. While promising the rate-guaranteed service to both primary and secondary users, we aim to maximize the primary rate. We develop an iterative algorithm that obtains the optimal offline power policies for primary and secondary users. To acquire insights about the structure of the optimal solution, we examine specific scenarios. Furthermore, we investigate the effects of the secondary rate constraint and finite battery on the primary rate and the probability of cooperation. We show that the joint information and energy cooperation increases the chances of cooperation and achieves significant rate gains over only information cooperation.  
\end{abstract}\vspace*{-1mm}

\section{Introduction}
Cognitive radio (CR)~\cite{goldsmith} has the potential to address the issue of inefficient use of scarce spectrum. The CR aims to improve the spectral efficiency by allowing spectrum sharing between licensed (primary) and unlicensed (secondary) users, without degrading the primary user's (PU's) performance. One such way is the cooperation between primary and secondary systems, where a secondary transmitter (ST) relays the primary transmitter's (PT's) information to the primary receiver (PR), and in turn, gains access to PU's spectrum to communicate with the secondary receiver (SR) as a reward. This cooperation not only enhances PU's quality-of-service (QoS), but also presents the secondary user (SU) better transmission opportunities. It serves as a better alternative to SU's opportunistic access (known as interweave mode~\cite{goldsmith}), as the latter needs SU to wait to find a spectrum hole through the dynamic process of spectrum sensing. Also, different from the underlay mode~\cite{goldsmith} that requires interference to PR below a threshold, the cooperation relaxes the constraint of low power transmission at ST.

In cooperative CR framework, most existing works (e.g., see \cite{devroye, sadek, jovicic,srini,simeone,pandhari:2009,hossain,active,cao,cao1,long}) have focused only on information cooperation between PU and SU. Consider the case of low energy availability with SU. In this case, if PT-PR channel is in deep fade and ST-PR channel is of good quality, neither PT can transmit its information nor ST can relay. Therefore, the information cooperation is not guaranteed if the relaying ST is energy-constrained. Energy harvesting (EH) is a promising solution to supply the perpetual energy to energy-constrained users. But, as harvesting energy from ambient sources like solar, wind, and vibration is random in nature, the availability of sufficient energy is not always assured, impeding the information cooperation. Thanks to the recent advances in energy transfer~\cite{karal,gurakan, yener}, the energy unavailability can be overcome by sharing energy between the nodes. Thus, expanding the cooperation to the energy level between PU and SU in addition to the information increases the probability of cooperation, improving the system's overall spectral efficiency. This motivates us to study the information and energy cooperation together in CR. 

The information cooperation in CR is initially studied in \cite{devroye, sadek, jovicic}, where SU has non-casual primary message knowledge, which it uses to eliminate PU interference by employing dirty paper coding. In \cite{cao,cao1}, an interference-free information cooperation is proposed using quadrature signaling. Authors in \cite{gurakan,yener} consider non-CR scenarios where a transmitter furnishes the relay with energy for information transfer, while authors in~\cite{rui} propose the joint information and energy cooperation in EH cellular networks. In CR with infinite battery, a recent work in~\cite{zheng} studies the joint information and energy cooperation between PU and SU in a single slot with an objective to provide best-effort QoS (does not assure the minimum QoS) for SU while guaranteeing minimum QoS to PU. However, when SU is entitled to provide real-time service requiring the minimum rate guarantee, provisioning the best-effort QoS may not be useful.

The main contributions of this paper are as follows:
\begin{itemize}
\item We address the joint information and energy cooperation in EH CR while providing rate-guaranteed service to both PU and SU. In addition, PU and SU desire to cooperate over a finite horizon, i.e., over multiple time-slots, under the finite battery constraint at ST. The SU helps PU relay the latter's information, while PU transfers energy to SU to facilitate the information cooperation.
\item In the proposed setup, we set the objective to maximize PU's rate, which is well-founded as PU supplies energy to the energy-constrained SU and thus seeks the \textit{best-effort} rate once the rate constraints for both users are satisfied. Since the proposed problem contains coupling variables, we decompose the original problem into subproblems and develop an iterative algorithm based on primal decomposition~\cite{palomar1} to obtain the offline optimal power policies for PU and SU.
\item To gain insights about the optimal solution, we consider specific scenarios corresponding to PT-PR and ST-PR links which contribute to the primary's rate. We show that when PT-PR link is better than ST-PR link, depending on the amount of the harvested energy by ST and network's channel gains, one-way energy cooperation from PU may expand to the virtual two-way energy cooperation between PU and SU. 
\end{itemize}   
\textit{Notation}: A bold-faced symbol (e.g., $\boldsymbol{P}$) denotes the vector of length $N$. The term $\sum_{i = a}^{b}(\cdot) = 0$ if $b < a$. The notation $[x]^+$ means $\max(0, x)$. The symbol $P^*$ denotes the optimal value, while $\boldsymbol{P}^*$ corresponds to the sequence of $P^*_is$ for $i = 1, \dotsc, N$. Throughout the paper, we assume $k = 1, \dotsc, N$.

\section{System Model and Cooperation Protocol}

\begin{figure}
\centering
\includegraphics[scale=0.33]{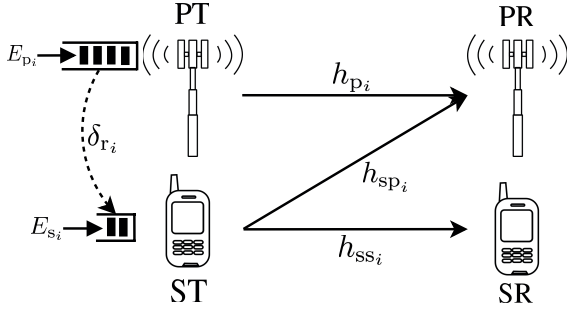}\vspace*{-3mm}
\caption{Information and energy cooperation in slot $i$.}
\label{fig:11}\vspace*{-5mm}
\end{figure}

We consider that PT and ST are energy harvesting nodes, and have no other conventional energy source. The ST has a battery of finite capacity $B_{\max}$ to store the incoming energy; whereas PT harvests energy at a higher rate than that of ST and has a large battery of sufficient capacity\footnote{This is generally the case when PT is a primary base-station, equipped with sophisticated energy harvesting devices.}. As shown in Fig.~\ref{fig:11}, to realize the cooperation, ST aids PT by relaying its traffic to PR, and in turn, receives an opportunity to transmit its own data to SR. Without loss of generality, we consider each slot duration to be unity. Hence, the terms energy and power can be used interchangeably. We focus on the case where PT and ST are located in close proximity to each other, and thus, the time required by ST to learn PT's message can be neglected~\cite{devroye, sadek, jovicic,srini}. We assume that PT and ST always have data to transmit. To allow concurrent transmissions by PT and ST without mutual interference, we employ the following cooperation framework as in~\cite{cao,cao1}. The PT and ST transmit on orthogonal channels by exploiting quadrature components of quadrature phase shift keying (QPSK). That is, PT transmits data to PR using binary phase shift keying, which is in-phase component of QPSK; ST uses QPSK for transmission, where it relays PT's traffic using in-phase component of QPSK (I-channel) and transmits its own data using quadrature-phase component (Q-channel). The PR can coherently combine the received signals from PT and ST.

In slot $i$, let $g_{\mathrm{p}_i}$, $g_{\mathrm{sp}_i}$, and $g_{\mathrm{ss}_i}$ denote Rayleigh channel power gains of PT-PR, ST-PR, and ST-SR links, respectively. The channel remains static in a slot, but changes independently over slots. The energy harvesting process at both PU and SU is assumed to be stationary and ergodic~\cite{parag}. We use the Bernoulli model for illustration, which is as follows: At the beginning of slot $i$, PU harvests energy $E_{\mathrm{p}_{i}}$, which is equal to $E_{\mathrm{p}}$ with probability $\theta_{\mathrm{p}}$ and zero otherwise; while SU harvests energy $E_{\mathrm{s}_{i}}$ which is equal to $E_{\mathrm{s}}$ with probability $\theta_{\mathrm{s}}$ and zero otherwise. The PT transmits with power $P_{\mathrm{d}_i}$ on PT-PR link and transfers energy $\delta_{\mathrm{r}_i}$ to ST. Note that PT transmits data and energy over orthogonal channels to ST~\cite{gurakan,yener}. The ST receives an energy $\alpha \delta_{\mathrm{r}_i}$, where $ \alpha $ ($0 < \alpha \leq 1$) is the energy transfer efficiency. We assume that PT and ST have the perfect global channel state information~\cite{zheng}.

The PR employs maximal ratio combining (MRC) to combine the signals from PT and ST. Let $R_{\mathrm{p,c}_{i}}$ and $R_{\mathrm{p,nc}_{i}}$ denote the rates achieved by PU with and without cooperation in $i$th slot, respectively, which are as follows:\vspace*{-1mm}
\begin{equation}
{R_{\mathrm{p,c}_i}}  =   \ln\left( 1 + {h_{\mathrm{p}_i}} P_{\mathrm{d}_i} +  {h_{\mathrm{sp}_i}}{P_{\mathrm{sp}_i}}\right),\vspace*{-1mm}
\end{equation}
\begin{equation}
\hspace*{-15mm}     {R_{\mathrm{p,nc}_i}}  =  \ln\left( 1+ {h_{\mathrm{p}_i}}P'_{\mathrm{d}_i}\right),\vspace*{-1mm}
\end{equation}
where ${P_{\mathrm{sp}_i}}$ denotes the power with which ST performs relaying, $h_{\mathrm{p}_i} = g_{\mathrm{p}_i}/N_0$ and $h_{\mathrm{sp}_i} = g_{\mathrm{sp}_i}/N_0$ are the normalized channel power gains on PT-PR and ST-PR links, respectively, and $N_0$ is the additive white Gaussian noise (AWGN) power. The power $P'_{\mathrm{d}_i}$  on PT-PR link under no cooperation is obtained by maximizing the objective $\sum_{i=1}^{N}{R_{\mathrm{p,nc}_i}}$ subject to the constraint $\sum_{i=1}^{k}P'_{\mathrm{d}_i} \leq \sum_{i=1}^{k}E_{\mathrm{p}_i}$ for $\forall k$. The secondary rate $R_{\mathrm{s}_i}$ in slot $i$ is\vspace*{-1mm}
\begin{equation}
{R_{\mathrm{s}_i}} = \ln\left( 1 + {h_{\mathrm{ss}_i}}{P_{\mathrm{ss}_i}}\right),\vspace*{-1mm} 
\end{equation}
where ${P_{\mathrm{ss}_i}}$ is the power with which ST transmits to SR and $h_{\mathrm{ss}_i} = g_{\mathrm{ss}_i}/N_0$. We now define the constraints in our problem before formulating it.\\
\textit{Cooperation rate constraint}: The PU's average rate with cooperation over $N$ slots must be at least its average rate without cooperation, i.e.,\vspace*{-1mm}
\begin{equation}
\frac{1}{N}\sum_{i=1}^N {R_{\mathrm{p,c}_i}} \geq \frac{1}{N}\sum_{i=1}^N {R_{\mathrm{p,nc}}}_{i} = \bar{R}_\mathrm{p}.\vspace*{-1mm}
\end{equation}
\textit{Secondary rate constraint}: The SU should achieve minimum average rate $\bar{R}_{\mathrm{s}}$ over $N$ slots during cooperation. That is,\vspace*{-1mm}
\begin{equation}
\frac{1}{N}\sum_{i = 1}^{ N} \ln\left( 1 +{h_{\mathrm{ss}_i}}{P_{\mathrm{ss}_i}} \right) \geq \bar{R}_{\mathrm{s}}.\vspace*{-1mm}
\label{eq:con_rs}
\end{equation}
\textit{Energy neutrality constraints}: 
The energy spent by PT till any slot $k$ cannot exceed the total harvested energy till that slot. Thus, over a finite horizon of $N$ slots, we have\vspace*{-2mm}
\begin{equation}
\sum_{i = 1}^{k} \left({P_{\mathrm{d}_i}} + {\delta_{\mathrm{r}_i}}\right) \leq \sum_{i = 1}^{k} {E_{\mathrm{p}_{i}}}, \quad \forall k.\vspace*{-1mm}
\end{equation}
Similarly, for ST, we write the energy neutrality constraint as\vspace*{-1mm}
\begin{equation}
\sum_{i = 1}^{k} \left({P_{\mathrm{sp}_i}} + {P_{\mathrm{ss}_i}}\right) \leq \sum_{i = 1}^{k} \left({E_{\mathrm{s}_{i}}} +  \alpha \delta_{\mathrm{r}_i}\right), \quad \forall k.\vspace*{-1mm}
\end{equation}
\textit{Finite battery constraint}: The ST cannot store more energy than the finite capacity $B_{\max}$ of the battery. That is,\vspace*{-1mm}
\begin{equation}
\sum_{i = 1}^{k}\left({E_{\mathrm{s}_i}} +  \alpha  \delta_{\mathrm{r}_i}\right)  - \sum_{i = 1}^{k-1}\left({P_{\mathrm{sp}_i}} + {P_{\mathrm{ss}_i}}\right)  \leq B_{\max}, \quad \forall k.\vspace*{-1mm}
\end{equation}

\section{Problem Formulation and Optimal Solution}
The objective is to maximize PU's rate with cooperation over a finite horizon of $N$ slots, i.e., $\sum_{i=1}^{N}R_{\mathrm{p,c}_i} $. The entire problem is now formulated as follows:\vspace*{-2mm}
\begin{subequations}
\label{eq:prob1}
\begin{align}
& \underset{\boldsymbol{P_{\mathrm{sp}}},\boldsymbol{P_{\mathrm{ss}}}}{\underset{\boldsymbol{P_{\mathrm{d}}},\boldsymbol{\delta_{\mathrm{r}}},}{\max}}
& & \sum_{i=1}^N {R_{\mathrm{p,c}_i}}, \label{eq:obj} \\
&\mathrm{s.t.} & & \frac{1}{N} \sum_{i=1}^N {R_{\mathrm{p,c}_i}} \geq \bar{R}_\mathrm{p}, \label{eq:Rp_con} \\
& & &\frac{1}{N}\sum_{i = 1}^{ N} \ln\left( 1 + {h_{\mathrm{ss}_i}}{P_{\mathrm{ss}_i}}\right) \geq \bar{R}_{\mathrm{s}},\label{eq:Rs_con} \\
& & & \sum_{i=1}^{k}\left({P_{\mathrm{d}_i}} + {\delta_{\mathrm{r}_i}}\right) \leq \sum_{i = 1}^{k}{E_{\mathrm{p}_i}} \quad \forall k,\label{eq:enc_p}\\
& & &  \sum_{i = 1}^{k} \left({P_{\mathrm{sp}}}_{i} + {P_{\mathrm{ss}_i}}\right)  \leq  \sum_{i = 1} ^ {k} \left({E_{\mathrm{s}_i}} + \alpha \delta_{\mathrm{r}_i}\right) \hspace{-1mm}\quad  \forall k , \label{eq:enc_s}\\
& & &  \sum_{i = 1}^{k} \left(\!{E_{\mathrm{s}_i}} + \alpha \delta_{\mathrm{r}_i}\right)  - \!\sum_{i = 1}^{k-1}\!\left({P_{\mathrm{sp}_i}} + {P_{\mathrm{ss}_i}}\right)  \leq \! B_{\max}  \quad \!\!\forall k, \label{eq:fin_bmax}\\
& & & {P_{\mathrm{d}_i}}, {\delta_{\mathrm{r}_i}}, {P_{\mathrm{sp}_i}}, {P_{\mathrm{ss}_i}}  \geq 0, \hspace{2mm} \forall i. \label{eq:non-neg}
\end{align}
\end{subequations}\vspace*{-5mm}

\noindent Since the objective and constraints \eqref{eq:Rp_con}, \eqref{eq:Rs_con} are concave, and the constraints \eqref{eq:enc_p}-\eqref{eq:fin_bmax} are affine, the problem \eqref{eq:prob1} is convex. The feasible region $ \mathcal{F}$ of problem \eqref{eq:prob1} is defined by the constraints \eqref{eq:Rp_con} - \eqref{eq:non-neg}. Firstly, we propose the necessary conditions that our optimal solution must satisfy and then proceed to find the optimal solution.\vspace*{-2mm}  
\subsection{Optimality Conditions}
\begin{proposition}
The optimal power policy ${P^*_{\mathrm{ss}_i}}$ ($1 \leq i \leq N$) must meet $\bar{R}_{{\mathrm{s}}} = \frac{1}{N}\sum_{i = 1}^{ N} \ln\left( 1 + {h_{\mathrm{ss}_{i}}}{P_{\mathrm{ss}_{i}}^{*}}\right)$.
\label{prop:1}
\end{proposition}\vspace*{-2mm}
\begin{proof} Assume the constraint \eqref{eq:Rs_con} is satisfied with strict inequality. We can then reduce some ${P^*_{\mathrm{ss}_i}}$ without violating the constraint and increase the power to relay PU's  data; say, increasing some $P^*_{\mathrm{sp}_i}$, which improves PU's rate. This contradicts the assumption that ${P^*_{\mathrm{ss}_i}}$ is optimal.
\end{proof}\vspace*{-1mm}
\begin{proposition}
\label{prop:2}
The optimal power policy ${P^*_{\mathrm{d}_i}}$ and ${\delta^*_{\mathrm{r}_i}}$ must meet $\sum_{i = 1}^{N} \left({P^*_{\mathrm{d}_i}} + {\delta^*_{\mathrm{r}_i}}\right)  = \sum_{i = 1}^{N}{E_{\mathrm{p}_i}}$.
\end{proposition}\vspace*{-1mm}
\begin{proof} 
Suppose the constraint~\eqref{eq:enc_p} is satisfied with strict inequality. That is, PT is left with some unused energy at the end of $N$ slots. Then, we can increase either some ${P^*_{\mathrm{d}_i}}$, or some ${\delta^*_{\mathrm{r}_i}}$ which contributes to ${P^*_{\mathrm{sp}_i}}$ without violating the constraint, achieving a higher PU rate. This contradicts the assumption that ${P^*_{\mathrm{d}_i}}$ and ${\delta^*_{\mathrm{r}_i}}$ are optimal.
\end{proof}\vspace*{-1mm}
\begin{proposition}
\label{prop:3}
The optimal power policy ${P^*_{\mathrm{ss}_i}}$, ${P^*_{\mathrm{sp}_i}}$, and $\delta^*_{\mathrm{r}_i}$ must meet $\sum_{i = 1}^{N} \left({P^*_{\mathrm{sp}}}_{i} + {P^*_{\mathrm{ss}_i}}\right)  = \sum_{i = 1}^{N} \left({E_{\mathrm{s}_i}} + \alpha \delta^*_{\mathrm{r}_i}\right)$.
\end{proposition}\vspace*{-1mm}
\begin{proof}
We omit this proof as it can be obtained based on the similar argument given in the proof for Proposition~\ref{prop:2}.
\end{proof}\vspace*{-2mm}

\subsection{Optimal Solution}\vspace*{-1mm}
We rewrite the objective in \eqref{eq:prob1} by incorporating \eqref{eq:Rp_con} in the objective as  $\sum_{i=1}^N {R_{\mathrm{p,c}}}_{i} - N\bar{R}_{\mathrm{p}}$, which is non-negative in the feasible region $\mathcal{F}$. Then, the Lagrangian of the problem \eqref{eq:prob1} is\vspace*{-2mm}
\begin{align}
&\mathcal{L} =  \Big( \sum_{i = 1}^{N}  \ln\left( 1 + {h_{\mathrm{p}_i}} P_{\mathrm{d}_i} + {h_{\mathrm{sp}_i}}{P_{\mathrm{sp}_i}} \right) - N\bar{ R}_{\mathrm{p}} \Big) \nonumber\\
& - \lambda \! \left(N\bar{R}_{\mathrm{s}} - \sum_{i = 1}^{ N} \ln\left( 1 + {h_{\mathrm{ss}_i}}{P_{\mathrm{ss}_i}}\right) \right)  \nonumber \\
&- \sum_{k=1}^{N}\mu_k\left(\sum_{i=1}^{k}\left({P_{\mathrm{d}_i}} + {\delta_{\mathrm{r}_i}}-{E_{\mathrm{p}_i}}\right)\right)\nonumber\\
&- \sum_{k = 1} ^{N}\gamma_{k} \left(\sum_{i = 1}^{k} \left({P_{\mathrm{sp}_i}} + {P_{\mathrm{ss}_i}} -  {E_{\mathrm{s}_i}} - \alpha   \delta_{\mathrm{r}_i}\right) \right) \nonumber \\
&- \sum_{k = 1} ^{N} \gamma'_{k} \left( \sum_{i = 1}^{k} \left({E_{\mathrm{s}_i}} + \alpha  \delta_{\mathrm{r}_i}\right)  - \sum_{i = 1}^{k-1}\left({P_{\mathrm{sp}_i}}+ {P_{\mathrm{ss}_i}}\right)  - B_{\max}\right) \nonumber \\
&+ \sum_{k=1}^{N}\tau_{1,k}{P_{\mathrm{d}_k}} + \sum_{k=1}^{N}\tau_{2,k}{\delta_{\mathrm{r}_k}}  + \sum_{k=1}^{N}\tau_{3,k}{P_{\mathrm{sp}_k}} + \sum_{k=1}^{N}\tau_{4,k}{P_{\mathrm{ss}_k}},
\label{eq:opt2}
\end{align}\vspace*{-4mm} 

\noindent where  $\lambda$, ${\boldsymbol{\mu}}$, ${\boldsymbol{\gamma}}$, ${\boldsymbol{\gamma'}}$, $\boldsymbol{\tau_{1}}$, $\boldsymbol{\tau_{2}}$, $\boldsymbol{\tau_{3}}$, and $\boldsymbol{\tau_{4}}$ are dual variables. The Karush-Kuhn-Tucker (KKT) stationarity conditions are\vspace*{-5mm}

\begin{align}
 \frac{h_{\mathrm{p}_i}}{1+{h_{\mathrm{p}_i}} P^*_{\mathrm{d}_i} + {h_{\mathrm{sp}_i}}{P^*_{\mathrm{sp}_i}}} - \sum_{k=i}^{N} \mu^*_k +\tau^*_{1,i} = 0, \label{KKT1} &\\ 
 -\sum_{k=i}^{N}\mu^*_k +\alpha\sum_{k=i}^{N} \gamma^*_k -\alpha\sum_{k=i}^{N} \gamma'^*_k + \tau^*_{2,i}=0,\label{KKT2} & \\
\frac{h_{\mathrm{sp}_i}}{1+{h_{\mathrm{p}_i}} P^*_{\mathrm{d}_i} + {h_{\mathrm{sp}_i}}{P^*_{\mathrm{sp}_i}}} - \!\sum_{k=i}^{N} \gamma^*_k +\!\sum_{k=i+1}^{N}\! \gamma'^*_k+\tau^*_{3,i} = 0, \label{KKT3}& \\
\frac{\lambda^* h_{\mathrm{ss}_i}}{1+ h_{\mathrm{ss}_i}P^*_{\mathrm{ss}_i}} - \sum_{k=i}^{N} \gamma^*_k +\sum_{k=i+1}^{N} \gamma'^*_k+\tau^*_{4,i} = 0,& \label{KKT4}
\end{align}\vspace*{-3mm}

\noindent and the complementary slackness conditions are as follows:\vspace*{-2mm}
\begin{align}
\lambda^* \left(N\bar{R}_{\mathrm{s}} - \sum_{i = 1}^{ N} \ln\left( 1 + {h_{\mathrm{ss}_i}}{P^*_{\mathrm{ss}_i}}\right) \right)  = 0,&  \\
\mu^*_k\sum_{i=1}^{k}\left({P^*_{\mathrm{d}_i}} + {\delta^*_{\mathrm{r}_i}}-{E_{\mathrm{p}_i}}\right) = 0,&\\
\gamma^*_{k}\sum_{i=1}^{k} \left({P^*_{\mathrm{sp}}}_{i} + {P^*_{\mathrm{ss}}}_{i} - {E_{\mathrm{s}}}_{i} -  \alpha \delta^*_{\mathrm{r}_i} \right)= 0,  &\\
\!\gamma'^*_{k} \! \left(\! \sum_{i = 1}^{k} \left({E_{\mathrm{s}_i}} + \alpha  \delta^*_{\mathrm{r}_i}\right)  - \!\sum_{i = 1}^{k-1}\left({P^*_{\mathrm{sp}_i}}+ {P^*_{\mathrm{ss}_i}}\!\right)  - B_{\max}\!\right) \!=\! 0,& \label{eq:slack_bmax} \\
\tau^*_{1,k}{P^*_{\mathrm{d}_k}}= \tau^*_{2,k}{\delta^*_{\mathrm{r}_k}}=\tau^*_{3,k}{P^*_{\mathrm{sp}_k}}=\tau^*_{4,k}{P^*_{\mathrm{ss}_k}} = 0 \label{eq:slack_power}&
\end{align}\vspace*{-5mm}

\noindent for all $k$. For the ease of computation, we neglect the dual variables $\boldsymbol{\tau}_{1}, \boldsymbol{\tau}_{2}, \boldsymbol{\tau}_{3}$, and $\boldsymbol{\tau}_{4}$ associated with the non-negativity of power vectors. Rather, we incorporate the non-negativity by projecting the powers onto the positive orthant and use them wherever necessary. From \eqref{KKT1}-\eqref{KKT4}, we obtain\vspace*{-2mm}

{{\small
\begin{equation}
\hspace*{-23mm} P^{*}_{\mathrm{d}_i}  =  \left[\frac{1}{\sum_{k = i}^N\mu^{*}_k }-\frac{1}{h_{\mathrm{p}_i}} - \frac{h_{\mathrm{sp}_i}}{h_{\mathrm{p}_i}}P^{*}_{\mathrm{sp}_i}\right]^{+},
\label{eq:sol1}\vspace*{-1mm}
\end{equation}}}\vspace*{-4mm}

{{\small
\begin{equation}
P^{*}_{\mathrm{sp}_i}  =  \left[\frac{1}{\sum_{k = i}^N\gamma^{*}_k - \sum_{k = i+1}^N\gamma'^{*}_k}-\frac{1}{h_{\mathrm{sp}_i}} - \frac{h_{\mathrm{p}_i}}{h_{\mathrm{sp}_i}}P^{*}_{\mathrm{d}_i}\right]^{+}\!\!,
\label{eq:sol2}\vspace*{-1mm}
\end{equation}}}

\vspace*{-7mm}
\begin{equation}
\hspace*{-15mm} P^{*}_{\mathrm{ss}_i} = \left[\frac{\lambda}{\sum_{k=i}^{N} \gamma^{*}_k -\sum_{k=i+1}^{N} \gamma'^{*}_k}-\frac{1}{h_{\mathrm{ss}_i}}\right]^{+},
\label{eq:sol3}\vspace*{-1mm}
\end{equation}
\begin{equation}
\frac{\alpha h_{\mathrm{sp}_i} - h_{\mathrm{p}_i} }{1 + h_{\mathrm{p}_i}P^{*}_{\mathrm{d}_i}+h_{\mathrm{sp}_i}P^{*}_{\mathrm{sp}_i}}= \tau^{*}_{1,i}+\alpha \gamma'^{*}_i - \tau^{*}_{2,i} - \alpha \tau^{*}_{3,i}.
\label{eq:sol4}\vspace*{-1mm}
\end{equation}
Note that, given the strict concave nature of rate constraints at PT and ST, the associated optimal power vectors (${\boldsymbol{P}^{*}_{\mathrm{d}}}, {\boldsymbol{P}^{*}_{\mathrm{sp}}},{\boldsymbol{P}^{*}_{\mathrm{ss}}}$) are unique; whereas, ${\boldsymbol{\delta}^{*}_{\mathrm{r}}}$ can have multiple possible values.
Due to the affine nature of constraints \eqref{eq:enc_p}-\eqref{eq:fin_bmax} associated with $\boldsymbol{\delta}^{*}_{\mathrm{r}}$, the expression to evaluate $\boldsymbol{\delta}^{*}_{\mathrm{r}}$ cannot be found.
Also, $\boldsymbol{P}^{*}_{\mathrm{d}}$ and $\boldsymbol{P}^{*}_{\mathrm{sp}}$ are dependent on each other as seen from~\eqref{eq:sol1} and \eqref{eq:sol2}. Thus, to compute the optimal power vectors based on \eqref{eq:sol1}-\eqref{eq:sol3}, we employ an iterative algorithm, which is explained below.\vspace*{-1mm}

\subsection{Iterative Algorithm to Compute Optimal Solution}

From \eqref{eq:prob1}, we observe that the power transfer variable $\boldsymbol{\delta}_{\mathrm{r}}$ couples the energy neutrality constraints at PT and ST given by \eqref{eq:enc_p}, \eqref{eq:enc_s} and the finite battery constraint at ST in  \eqref{eq:fin_bmax}. Also, the constraint \eqref{eq:Rp_con} couples the powers  $\boldsymbol{P}_{\mathrm{d}}$ and $\boldsymbol{P}_{\mathrm{sp}}$. Now, to decouple the variables and find the optimal solution, we propose the following approach. Firstly, to decouple \eqref{eq:enc_p} and \eqref{eq:enc_s}-\eqref{eq:fin_bmax}, we perform the primal decomposition~\cite{palomar1} by fixing the coupling variable $\boldsymbol{\delta}_{\mathrm{r}}$. Then, we divide the problem \eqref{eq:prob1} into three layers. In each iteration, Layer $\mathrm{1}$ solves the power allocation for $\boldsymbol{P}_{\mathrm{d}}$, while Layer $\mathrm{2}$ solves the power allocation for $(\boldsymbol{P}_{\mathrm{sp}}, \boldsymbol{P}_{\mathrm{ss}})$, and finally, Layer $\mathrm{3}$ updates $\boldsymbol{\delta}_{\mathrm{r}}$. Note that Layers $\mathrm{1}$ and $\mathrm{2}$ decouple the powers $\boldsymbol{P}_{\mathrm{d}}$ and $\boldsymbol{P}_{\mathrm{sp}}$.  This three-layer problem is solved in an iterative manner until all optimization variables converge. 
To begin with, we initialize the primal variables $\boldsymbol{\delta}_{\mathrm{r}}, \boldsymbol{P}_{\mathrm{sp}}, \boldsymbol{P}_{\mathrm{ss}}$,  and dual variables  $\lambda, \boldsymbol{\mu}, \boldsymbol{\gamma}, \boldsymbol{\gamma'}$.
Below, we explain the sub-problems in each layer.

\textit{Layer 1}: The sub-problem to solve for the power vector $\boldsymbol{P}_{\mathrm{d}}$ on the direct link of PT is \vspace{-1mm}
\begin{align}
& \underset{\boldsymbol{P}_{\mathrm{d}}}{\max}
  \hspace{6mm} \sum_{i=1}^N {R_{\mathrm{p,c}_i}} -N \bar{R}_{\mathrm{p}}\nonumber \\  &   \mathrm{s.t.}  \hspace{8mm} \eqref{eq:enc_p}, P_{\mathrm{d}_i} \geq 0 \hspace{2mm} \forall i.
\label{eq:lag1}
\end{align}\vspace*{-5mm}

\noindent For a given $\boldsymbol{\delta}_{\mathrm{r}}$, $\boldsymbol{P}_{\mathrm{sp}}$, and $\boldsymbol{\mu}$, $\boldsymbol{P}_{\mathrm{d}}$ is evaluated using \eqref{eq:sol1}. The  dual problem of \eqref{eq:lag1}  is given by 
$\underset{ \boldsymbol{\mu} \geq 0} \min \hspace{2mm} \underset{\boldsymbol{P}_{\mathrm{d}} \in \mathcal{F}} \max \hspace{2mm} \mathcal{L}_{1}$,
where $\mathcal{L}_{1}$ is the corresponding Lagrangian function. Since the dual function $ \underset{\boldsymbol{P}_{\mathrm{d}} \in \mathcal{F}} \max \hspace{2mm} \mathcal{L}_{1}$ is differentiable, the dual variable $\boldsymbol{\mu}$ that minimizes the dual problem in \eqref{eq:lag1} is found using gradient method as \vspace{-2mm}
\begin{equation*}
\mu_{k} = \left[\mu_{k} + s  \sum_{i=1}^{k} \left({P_{\mathrm{d}_i}} + {\delta_{\mathrm{r}_i}}-{E_{\mathrm{p}_i}} \right) \right]^{+}, \forall k,\vspace*{-1mm}
\end{equation*}
where $s$ denotes the positive step size chosen to satisfy the diminishing step size rule~\cite{bertesekas}.

\textit{Layer 2}: The sub-problem for jointly solving the power allocation $(\boldsymbol{P}_{\mathrm{sp}}, \boldsymbol{P}_{\mathrm{ss}})$ is \vspace{-2mm}
\begin{align}
& \underset{\boldsymbol{P}_{\mathrm{sp}}, \boldsymbol{P}_{\mathrm{ss}}}{\max}
& & \sum_{i=1}^N {R_{\mathrm{p,c}_i}} -N \bar{R}_{\mathrm{p}} \nonumber \\
&\mathrm{s.t.} & & \eqref{eq:Rs_con}, \eqref{eq:enc_s}, \eqref{eq:fin_bmax},  P_{\mathrm{sp}_i}, P_{\mathrm{ss}_i} \geq 0 \hspace{2mm}\forall i.
\label{eq:lag2}
\end{align}\vspace*{-5mm}

\noindent Using the results obtained in \eqref{eq:sol2} and \eqref{eq:sol3}, we compute $(\boldsymbol{P}_{\mathrm{sp}}, \boldsymbol{P}_{\mathrm{ss}})$ for given $\boldsymbol{P}_{\mathrm{d}}$,  $\boldsymbol{\delta}_{\mathrm{r}}$,  and $\lambda, \boldsymbol{\gamma}, \boldsymbol{\gamma'}$.
As in Layer $\mathrm{1}$, the dual variables minimizing the dual problem of \eqref{eq:lag2} given by $\underset{\lambda, \boldsymbol{\gamma}, \boldsymbol{\gamma'}, \geq 0} \min \hspace{2mm} \underset{\boldsymbol{P}_{\mathrm{sp}},\boldsymbol{P}_{\mathrm{ss}} \in \mathcal{F}} \max \hspace{2mm} \mathcal{L}_{2} $, are updated using their gradients as\vspace*{-3mm}

{{\small
\begin{align*} \vspace{-3mm}
\lambda & = \left[\lambda + s \left(N\bar{R}_{\mathrm{s}} - \sum_{i = 1}^{ N} \ln\left( 1 + {h_{\mathrm{ss}_i}}{P_{\mathrm{ss}_i}}\right) \right) \right]^{+}, \\
\gamma_{k} & = \left[ \gamma_{k} + s \left( \sum_{i=1}^{k} \left({P_{\mathrm{sp}}}_{i} + {P_{\mathrm{ss}}}_{i} - {E_{\mathrm{s}}}_{i} -  \alpha  \delta_{\mathrm{r}_i}\right)  \right) \right]^{+}, \\
\gamma'_{k} & = \left[\! \gamma'_{k} + s \left(\! \sum_{i = 1}^{k}\! \left({E_{\mathrm{s}_i}} \!+\! \alpha  \delta_{\mathrm{r}_i}\right)  - \!\sum_{i = 1}^{k-1}\!\left({P_{\mathrm{sp}_i}}+ {P_{\mathrm{ss}_i}}\!\right)  - B_{\max} \! \right)\! \right] ^{+}\!\!\!
\end{align*}}}\vspace*{-3mm}

\noindent for all $k$, and $\mathcal{L}_{2}$ is the Lagrangian of \eqref{eq:lag2}. The computation of the set of primal, dual variables $(\boldsymbol{P}_{\mathrm{d}}, \boldsymbol{\mu})$, and $(\boldsymbol{P}_{\mathrm{sp}},\boldsymbol{P}_{\mathrm{ss}},\lambda, \boldsymbol{\mu}, \boldsymbol{\gamma}, \boldsymbol{\gamma'})$ is done recursively in their corresponding layers $\mathrm{1}$ and $\mathrm{2}$ until they converge to a predetermined accuracy.

\textit{Layer 3}: In Layer $\mathrm{3}$, the primal variable $\boldsymbol{\delta}_{\mathrm{r}}$ is updated using sub-gradient method as\vspace*{-2mm}
\begin{equation*}
\delta_{\mathrm{r}_i} = \left[\delta_{\mathrm{r}_i} - s \left(\sum_{j = i}^{N}\left(\mu_{j} -\alpha \gamma_{j} +\alpha \gamma'_{j}\right)\right) \right]^+ \forall i. 
\end{equation*}
Since each layer involves solving a convex problem, this three-layer iterative approach is guaranteed to converge to the optimal solution~\cite{bertesekas}.\vspace*{-1mm}

\section{Specific Scenarios}\vspace*{-1mm}
\label{spe}
In this section, we provide specific scenarios to gain insights about the joint information and energy cooperation protocol. The primary rate mainly depends on PT-PR and ST-PR links, characterized by channel power gains $h_{\mathrm{p}_i}$ and $h_{\mathrm{sp}_i}$, respectively. Unlike the only direct link transmission, i.e., via PT-PR link, PT can achieve user diversity with the help of ST and through ST-PR link. Hence, it is important to investigate the effects of both links on the optimal solution.\vspace*{-1mm}
\subsection{The case when $h_{\mathrm{p}_i} >  h_{\mathrm{sp}_i}$}
\label{sec:spec}
Given ST has the knowledge about the primary message, it looks natural for PT to allocate the power for its transmission on the link chosen from PT-PR and ST-PR links with the better channel power gain. That is, when $h_{\mathrm{p}_i} > h_{\mathrm{sp}_i}$, it may appear that $P_{\mathrm{sp}_i}$ is zero. However, this is not always true, which we show through the following proposition.\vspace*{-1mm}
\begin{proposition}
When $h_{\mathrm{p}_i} > h_{\mathrm{sp}_i}$, ${\delta^*_{\mathrm{r}_i}}$ and ${P^*_{\mathrm{sp}_i}}$ cannot be non-zero simultaneously.
\label{prop:4}
\end{proposition}\vspace*{-1mm}
\begin{proof}
When $h_{\mathrm{p}_i} > h_{\mathrm{sp}_i}$, from \eqref{eq:sol4}, we observe that the left side term of the equation is negative since $0 < \alpha \leq 1$, and $\left(1 + h_{\mathrm{p}_i}P^*_{\mathrm{d}_i}+h_{\mathrm{sp}_i}P^*_{\mathrm{sp}_i}\right) > 0 $. Thus, the right side term of \eqref{eq:sol4} is also negative, making $(\tau^*_{2,i} + \alpha \tau^*_{3,i}) > (\tau^*_{1,i}+\alpha \gamma'^*_i)$. 

Suppose ${\delta^*_{\mathrm{r}_i}}$ and ${P^*_{\mathrm{sp}_i}}$ are non-zero in slot $i$. Then, from the complementary slackness condition in \eqref{eq:slack_power}, we have $\tau^*_{2,i} = \tau^*_{3,i} = 0$. This leads to $0 > (\tau^*_{1,i}+\alpha \gamma'^*_i)$, which is impossible as $\tau^*_{1,i} \geq 0$ and $\gamma'^*_i \geq 0$, arriving at a contradiction.
\end{proof}
Note that Proposition~\ref{prop:4} holds for both finite and infinite battery cases. We further elaborate Proposition~\ref{prop:4} intuitively as follows. In some slot $i$ with $h_{\mathrm{p}_i} > h_{\mathrm{sp}_i}$, PT is unwilling to spend power on ST-PR link, which is weaker than PT-PR link. Given this, PT has no incentive to transfer energy to ST, making ${\delta^*_{\mathrm{r}_i}} = 0$. Then, ${P^*_{\mathrm{sp}_i}} > 0$ means that ST relays PT's message with the energy left with it after satisfying its rate constraint $\bar{R}_{\mathrm{s}}$. On the other hand, the case ${\delta^*_{\mathrm{r}_i}} > 0$ and ${P^*_{\mathrm{sp}_i}} = 0$ occurs if ST is \textit{energy-depleted}. That is, PT feeds ST ${\delta^*_{\mathrm{r}_i}}$ amount of energy in that slot which either contributes towards satisfying SU's rate constraint or is to be utilized in future slots. 

\subsection{The case when $h_{\mathrm{p}_i} < \alpha h_{\mathrm{sp}_i}$}
\begin{proposition}
\label{prop:5}
When $h_{\mathrm{p}_i} < \alpha h_{\mathrm{sp}_i}$, \vspace*{-1mm}
\begin{equation}
\label{eq:pdf}
{P^*_{\mathrm{d}_i}}  \left\{
  \begin{array}{l l}
    = 0, & \quad E_{i} < B_{\max} \\
    \geq 0, & \quad E_{i} = B_{\max},
  \end{array} \right.\vspace*{-1mm}
\end{equation}
where $E_{i} =  \sum_{j = 1}^{i} \left({E_{\mathrm{s}_j}} + \alpha \delta^*_{\mathrm{r}_j}\right)  - \sum_{j = 1}^{i-1}\left({P^*_{\mathrm{sp}_j}} + {P^*_{\mathrm{ss}_j}}\right)$ is the energy available with ST at the start of slot $i$.
\end{proposition}\vspace*{-2mm}
\begin{proof}
When $h_{\mathrm{p}_i} < \alpha h_{\mathrm{sp}_i}$, from \eqref{eq:sol4}, we see that the left side term of the equation is positive, implying the right side term of \eqref{eq:sol4} is also positive, making $(\tau^*_{1,i}+\alpha \gamma'^*_i) > (\tau^*_{2,i} + \alpha \tau^*_{3,i})$. 

For $E_{i} < B_{\max}$, from \eqref{eq:slack_bmax}, we have $\gamma'^*_{i} = 0$. Thus, if ${P^*_{\mathrm{d}_i}} > 0$, $\tau^*_{1,i} = 0$ leading to $(\tau^*_{2,i} + \alpha \tau^*_{3,i})< 0$, which is impossible since $\tau^*_{2,i} \geq 0$ and $\tau^*_{3,i} \geq 0$. For $E_{i} = B_{\max}$, $\gamma'^*_{i} > 0$ from \eqref{eq:slack_bmax}.  Thus, ${P^*_{\mathrm{d}_i}}$ can be either non-zero with $\alpha \gamma'^*_i > \tau^*_{2,i} + \alpha \tau^*_{3,i}$ or zero with $\tau^*_{1,i}+\alpha \gamma'^*_i > \tau^*_{2,i} + \alpha \tau^*_{3,i}$.
\end{proof}
In some slot $i$, let us first consider $E_{i} < B_{\max}$. Here, PT is willing to spend its transmission power only on ST-PR link as $h_{\mathrm{p}_i} < \alpha h_{\mathrm{sp}_i}$. Thus, PT will transfer energy $\delta^*_{\mathrm{r}_i}$ to ST as long as $E_{i} < B_{\max}$, keeping ${P^*_{\mathrm{d}_i}} = 0$. However, with $E_{i} = B_{\max}$, ST cannot accommodate additional energy from PT, which forces PT to either transmit on PT-PR link or store the energy for the transmission in future slots, depending on future channels gains on PT-PR and ST-PR links.

\noindent \textbf{Numerical Example}: We provide a numerical example to illustrate
Propositions~\ref{prop:4} and \ref{prop:5}. Let $N = 5$, $\boldsymbol{h_\mathrm{p}} = [0.0191~0.0080~0.0036~0.0024~0.0119]$, $\boldsymbol{h_\mathrm{sp}} = [0.0065~0.0074~0.0194~0.0256~0.0067]$, $\boldsymbol{h_{\mathrm{ss}}} =$ $[0.0027~0.0140~0.0164~0.0201~0.0010]$, $\boldsymbol{E_{\mathrm{s}}} = [0~0~1~0~1]$, $\boldsymbol{E_{\mathrm{p}}} = [7~0~0~7~0]$, $B_{\mathrm{max}} = 3.5$ J, $\alpha = 1$, $N_0 = 0~\mathrm{dBm}$, $\bar{R}_{\mathrm{s}} =$ $\mathrm{0.5}~\mathrm{nats/slot/Hz}$. PU's sum-rate in cooperation scheme $R_{\mathrm{p,c}} = \mathrm{ 3.6914~nats/slot/Hz}$, and $R_{\mathrm{p,nc}} =\mathrm{ 3.0073~nats/slot/Hz}$ without cooperation. The optimal power allocation is given in Table~\ref{tab:2}.\vspace*{-2mm}

\begin{table}[h]
\centering
\caption{Optimal Solution to Numerical Example}\vspace*{-2mm}
\begin{tabular}{| l | l | l | l | l | l|}
\hline
    Slot & 1 & 2 & 3 &  4 & 5 \\ \hline
    $P_{\mathrm{ss}_i}^{*}$  &   0.0000  &  0.2249  &  0.2354  &  0.3165  &  0.0000 \\ \hline
    $P_{\mathrm{sp}_i}^{*}$  &  0.0000  &  0.0000  &  2.5014  &  3.1835  &  1.0000  \\ \hline
    $P_{\mathrm{d}_i}^{*}$ &  2.5555  &  2.4828   & 0.0000   & 0.0000   & 3.5000 \\ \hline
    $\delta_{\mathrm{r}_i}^{*}$  &   0.7216  &  0.4908 &  0.7493  &  3.5000 &   0.0000 \\ \hline 
    $E_{i}$  & 0.7216  & 1.2124  &  2.7368   & 3.5000  &  1.0000  \\  \hline 
\end{tabular}
\label{tab:2}
\end{table}\vspace*{-3mm}
We observe that PT-PR link is better than ST-PR link in the first, second, and last slots, where we have ${ {\delta^*_{\mathrm{r}_1}}, \delta^*_{\mathrm{r}_2}} > 0, {\delta^*_{\mathrm{r}_5}} = 0 $ and ${P^*_{\mathrm{sp}_1}}, {P^*_{\mathrm{sp}_2}} = 0, {P^*_{\mathrm{sp}_5}} > 0$ which is in agreement with Proposition~\ref{prop:4}. Also, we note that, energy transfers in the first two slots contribute towards ${P^*_{\mathrm{ss}_2}}$, i.e., it helps SU to achieve its rate constraint as ST has not harvested enough energy and is energy-depleted. In the last slot, we have ${\delta^*_{\mathrm{r}_5}} = 0$ and ${P^*_{\mathrm{sp}_5}} = 1$, implying ST has spent its harvested energy on the relaying link since its rate constraint is satisfied. This is also in agreement with Proposition~\ref{prop:3}, which says that no energy should be left with ST at the end of the horizon under the optimal solution. This could also be seen as a virtual two-way energy cooperation between PU and SU to maximize the primary rate given the rate constraints of both PU and SU are satisfied. Thus, the energy spent by PT satisfying SU's rate constraint in earlier slots is a wise investment by PU to improve its maximum achieved rate. Similarly, we can also validate the Proposition~\ref{prop:5} from Table~\ref{tab:2}.\vspace*{-1mm}

\section{Results and Discussions}\vspace*{-1mm}
We consider a scenario where the distances between PT-PR, ST-PR, and ST-SR are each 5~$\mathrm{m}$. The mean channel power gain of the channel between nodes $i$ and $j$ is given by $d_{ij}^{-\rho}$, where $d_{ij}$ is the distance between nodes $i$ and $j$ and $\rho$ is the path-loss exponent, which is assumed to be $\mathrm{2.7}$. ST is located close to PT at a distance of 0.5~$\mathrm{m}$. Thus, PT prefers the help from ST to relay the data and gains user diversity. Due to the close-proximity of PT and ST, ST can learn PT's message in negligible time~\cite{srini}. The noise power is 0~$\mathrm{dBm}$.  The probability of harvesting energy in a slot at PT and ST is $\theta_{\mathrm{p}} = \theta_{\mathrm{s}} = 0.5$. Unless otherwise stated, we obtain the results over $\mathrm{1000}$ channel and energy realizations.

\subsection{Effect of energy cooperation}
\begin{figure}
\centering
\includegraphics[scale=0.35]{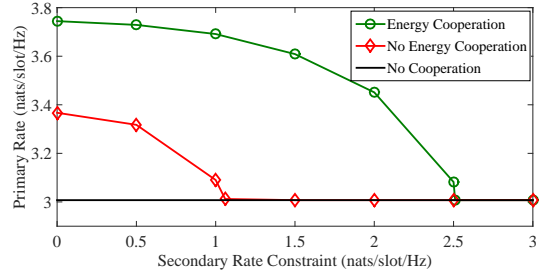}\vspace*{-3mm}
\caption{Rate regions with and without energy cooperation.}
\label{fig:2}\vspace*{-4mm}
\end{figure}

Fig.~\ref{fig:2} compares the rate regions for joint information and energy cooperation scheme to that of only information cooperation for a specific randomly chosen channel realization and system parameters given in the numerical example in Section \ref{spe}.
We observe that the joint information and energy cooperation expands the achievable rate region compared to that with only information cooperation. As seen from Fig.~\ref{fig:2}, in the case of only information cooperation, SU can impose the rate constraint ($\bar{R}_\mathrm{s}$) at the most $\mathrm{1.06}~\mathrm{nats/slot/Hz}$, beyond which increasing $\bar{R}_\mathrm{s}$ reduces the primary rate below the no cooperation rate, in turn, entering the infeasible region. Thus, the cooperation is not beneficial to PU any more and it pulls out of it achieving the no cooperation rate. On the other hand, enlarging the cooperation to energy level pushes the cooperation region to a higher $\bar{R}_\mathrm{s}$ ($\mathrm{2.51}~\mathrm{nats/slot/Hz}$ from $\mathrm{1.06}~\mathrm{nats/slot/Hz}$). Note that when there is no cooperation between PU and SU, PU transmits solely in the channel, and its rate is unaffected by $\bar{R}_\mathrm{s}$.
 
\subsection{Effect of secondary rate constraint }
\begin{figure}
\centering
\includegraphics[scale=0.35]{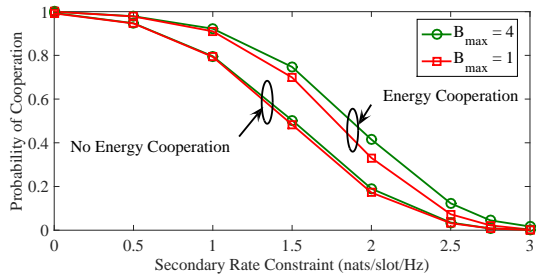}\vspace*{-3mm}
\caption{Probability of cooperation with and without energy cooperation, $N = 5$, $\alpha = 0.3$, $E_{\mathrm{p}} = 7$, $E_{\mathrm{s}} = 1$.}
\label{fig:1}\vspace*{-3mm}
\end{figure}

Fig.~\ref{fig:1} shows that the joint information and energy cooperation scheme between PU and SU increases the probability of cooperation compared to only information cooperation scheme, where the probability of cooperation is the ratio of number of channel and energy realizations that result in successful cooperation between PU and SU to the total number of channel and energy realizations considered in simulations. This is because, the energy cooperated by PU to the energy-constrained SU increases its lifetime and keeps the latter active in the network. Also, as SU imposes higher rate constraint $\bar{R}_{\mathrm{s}}$, the probability of cooperation in both cases reduces. This is due to the fact that, with higher $\bar{R}_{\mathrm{s}}$, it becomes more difficult to satisfy both primary and secondary rate constraints together given the energies harvested by PU and SU, making one of them to fall out of the cooperation. Another important observation is that the smaller ST's battery size ($B_{\mathrm{max}}$) reduces the probability of cooperation as less energy can be accommodated in SU's battery lowering the powers for relaying as well as its own transmission.

\begin{figure}
\centering
\includegraphics[scale=0.35]{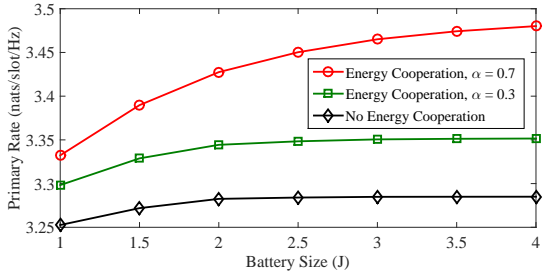}\vspace*{-3mm}
\caption{Effect of ST's finite battery size ($B_{\mathrm{max}}$), $N = 5$, $\bar{R}_{\mathrm{s}} = 0.5$ $\mathrm{nats/slot/Hz}$, $E_{\mathrm{p}} = 7$, $E_{\mathrm{s}} = 1$.}
\label{fig:4}\vspace*{-3mm}
\end{figure}

\subsection{Effect of battery size}

Fig.~\ref{fig:4} shows the effect of ST's battery size ($B_{\mathrm{max}}$), where the proposed joint information and energy cooperation scheme outperforms the only information cooperation scheme. We note that the increase in $B_{\mathrm{max}}$ improves the primary rate in both cases with and without energy cooperation as expected. In a slot, the lower $B_{\mathrm{max}}$ limits the energy transferred by PT to ST even if the latter has a better channel to PR. Thus, the gain achieved with energy cooperation over without energy cooperation scheme is low. But, with increase in $B_{\mathrm{max}}$, PT can cooperate more energy to the energy-constrained ST keeping the latter active for the cooperation and obtain significant rate gain over no energy cooperation scheme. If the battery size is sufficiently large to accommodate the harvested and transferred energy, there is a very limited additional advantage in making the battery size even bigger. Also, the higher energy transfer efficiency $\alpha$ makes more energy available to SU, increasing the probability of cooperation and thus the primary rate.

\section{Conclusions}
In this paper, we have studied the energy cooperation between energy harvesting primary and secondary users in addition to the information cooperation over a finite horizon with finite capacity battery. The joint information and energy cooperation expands the region of cooperation between primary and secondary users improving the chances of cooperation. This, in turn, increases the user diversity, through which the primary user achieves significant rate gains compared to only information cooperation while assuring the rate-guaranteed service to both users.

\bibliographystyle{ieeetr}
\bibliography{paper}
\end{document}